Powder X-Ray Diffraction Assisted Evolutionary Algorithm for Crystal Structure Prediction

Stefano Racioppi[a], Alberto Otero De la Roza[b], Samad Hajinazar[a], Eva Zurek[a]*

[a] Department of Chemistry, State University of New York at Buffalo, Buffalo, New York 14260-3000, USA.

[b] Departmento de Química Física y Analítica and MALTA Consolider Team, Facultad de Química, Universidad de Oviedo, 33006 Oviedo, Spain.

**Abstract**

Experimentally obtained X-ray diffraction (XRD) patterns can be difficult to solve, precluding the full characterization of materials, pharmaceuticals, and geological compounds. Herein, we propose a method based upon a multi-objective evolutionary search that uses both a structure's enthalpy and similarity to a reference XRD pattern (constituted by a list of peak positions and their intensities) to facilitate structure solution of inorganic systems. Because the similarity index is computed for locally optimized cells that are subsequently distorted to find the best match with the reference, this process transcends both computational (*e.g.* choice of theoretical method, and 0 K approximation) and experimental (*e.g.* external stimuli, and metastability) limitations. We illustrate how the proposed methodology can be employed to successfully uncover complex crystal structures by applying it to a range of test cases, including inorganic minerals, pure elements ramp-compressed to extreme conditions, and molecular crystals. The results demonstrate that our approach not only improves the accuracy of structure prediction but also significantly reduces the time required to achieve reliable solutions, thus providing a powerful tool for the advancement of materials science and related fields.

**Introduction**

The crystal structure of a compound is key for predicting and rationalizing its properties.[1–3] Therefore, crystal structure determination is one of the bedrocks upon which chemistry, materials science, physics, as well as earth and planetary science is based. Indeed, common to all of these fields is the need for characterizing the structure of the chemical system using various spectroscopies. While methods such as Raman, Infra-Red or Nuclear Magnetic Resonance spectroscopy provide information that can indirectly deduce the structural motifs present, only diffraction is directly related to the atomic positions. Various diffraction techniques are available, varying according to the scattering source (X-rays, neutrons or electrons) or from the nature of the sample (powder, single crystal or even liquid). Diffraction from single crystals is the gold standard, but in practice it can be difficult or impossible to obtain single crystals of adequate quality and size to achieve a reliable structure solution. Therefore, the possibility of obtaining structural information from microcrystalline powder-like samples becomes important.[4] However, unlike diffraction from single crystals, powder X-ray diffraction (PXRD) is typically not sensitive enough to provide information about the positions of light elements such as hydrogen and lithium (if combined with other heavier elements), nor can it differentiate between elements with similar mass numbers. Though neutron diffraction is sensitive to the location of light elements, it requires large samples. Thus, PXRD is the most commonly used tool to deduce the structural information of battery materials, superconductors, minerals found in the deep Earth, pharmaceutical drugs, and more.[1,5–7]

When a good quality PXRD pattern is in-hand, it is relatively easy to retrieve information on the size of the unit cell, but a structure solution with refinement of the atomic position remains, to date, a challenging procedure. The inherent limitation of PXRD lies in its projection of three-dimensional diffraction data onto a one-dimensional scale when measuring powder samples, often resulting in peak overlap.[4,8] To perform such refinements, various techniques[9–11] are available, from those developed by Rietveld[12] or Le Bail[13] to modellings based on reverse Monte Carlo,[4,14] genetic [15,16] or machine learning[17] algorithms. Nonetheless, crystal structure solution from PXRD data remains a grand challenge in crystallography. Further complicating structural characterization is the presence of mixed-phases, sample peaks that are obscured by ones originating from the experimental apparatus, noisy background of the diffractogram, and preferred orientation of the microcrystallites. These situations are common, for example, when compounds are synthesized for the first time, matter is compressed within diamond anvil cells or in dynamic (shock or ramp) compression experiments, or simply due to the morphology of the crystallites. Because of these difficulties, theoretical calculations have recently become useful tools to assist structure solution given a PXRD pattern.

Another strategy, popular especially in the high-pressure field, is based upon crystal structure prediction (CSP) algorithms, which aim to locate the most stable atomic configuration for a user-defined chemical composition at a given pressure and at zero temperature. Some of the most popular techniques include random or evolutionary searches, particle-swarm optimization, and Monte-Carlo or molecular dynamics based algorithms.[18] In the family of evolutionary algorithms, a fitness is assigned to each DFT-optimized structure, and this fitness is related to the structure's likelihood to be chosen as a parent for the next generation.[19] In a traditional evolutionary search, the fitness is obtained from the energy (or enthalpy) of the system relative to (a subset) of those that have been optimized. This fitness is crucial in driving the algorithm towards promising regions of the energy landscape in the search for thermodynamically stable structures.

These CSP-based methods have become invaluable tools for the characterization of unknown phases, facilitating structural determination from experimental data, and particularly from X-ray diffraction.[20] However, not all of the compounds that are predicted to be the most stable are necessarily those that are experimentally observed.[21] When compared to PXRD diffractograms, this discrepancy can be attributed to several factors, including the numerous approximations involved in the computations (*e.g.*, choice of level of theory, pseudopotential, and the neglect of finite temperature contributions), as well as variations in synthetic and experimental conditions [22]. As a result, achieving the closest match with the experimental data often necessitates screening many metastable phases, especially in the case of polymorphism, where the differences in energy between them may be minimal.[23,24] This laborious manual screening process carries the risk of overlooking the optimal matching structure amid the hundreds or even thousands of predicted structures.

To circumvent this challenge and potentially steer the structure search towards a better match, a guiding CSP algorithm, which employs both the experimental PXRD and the DFT-calculated energy and structure simultaneously, and in an equal footing could prove beneficial.[21] Perhaps the most intricate method proposed to date is the first-principle-assisted structure solution (FPASS) technique,[25] which combines DFT calculations with experimental XRD data and statistical symmetry information in a genetic algorithm for structure determination. Similar methods have followed during the years, retaining the philosophy of combining information gleaned from diffraction data (lattice parameter, symmetry, stoichiometry, etc.) to reduce the space of a structure search with DFT optimization[24,26] A similarity index calculated between experimental and simulated PXRD patterns has been exploited in particle swarm optimization (PSO)[27], utilizing a weighted cross-correlation function to re-evaluate the velocity of each structure in the crystal structure search. With this technique, the simulated phases that best match the experimental PXRD pattern will lead the PSO search. This methodology was shown to aid the prediction of the ground state phases of ZnO and $TiO_2$,[27] however the authors did not report if their discovery was accelerated compared to a standard energy-only-search, or if this algorithm could aid in the recognition of metastable phases. Non-CSP-based approaches have also been proposed, such as molecular dynamics simulations biased by experimental diffraction data [28]. For molecular organic crystals fast dispersion-corrected DFT optimizations have been proposed and used to improve the fit with the experimental PXRD patterns, instead.[29]

In the present study, we outline our approach for enhancing CSP by leveraging PXRD data through a synergistic combination of the XtalOpt evolutionary algorithm[19,30] and the variable-cell Gaussian powder-based similarity index (*VC-GPWDF*)[31] implemented in the critic2 program.[32] Confusingly, despite the label "similarity index", this and similar methods actually calculate the *dissimilarity* between two patterns. By using the multi-objective search capability embedded in XtalOpt,[33] we illustrate the CSP search is able to accelerate the structural recognition of both experimental and simulated PXRD patterns. This strategy goes beyond the aforementioned methodologies developed to assist CSP using crystallographic and diffraction information.[25,27] Specifically, it overcomes many of the challenges that result from the comparison of experimental diffraction data collected at finite temperature and pressure with *in-silico* patterns calculated for geometry-optimized structures at 0 K. Therefore, this technique becomes particularly advantageous when the reference PXRD diffractogram diverges from the computed ground state structure of a specific stoichiometry due to experimental conditions (pressure, hydrostaticity, temperature, etc.) or because of theoretical limitations; or for finding metastable phases.

**Results**

**Multi-objective search**

The foundation of our newly proposed technique is based upon multi-objective global optimization,[34] as implemented in the XtalOpt code version 13.0.[33] In this extension of the XtalOpt evolutionary algorithm, the fitness of an individual structure can be based upon multiple objectives, including a structure's energy or enthalpy, as well as other user-specified features. After locally relaxing a structure, XtalOpt automatically calls the external codes specified by the user to compute the desired target properties, whose values are employed in conjunction with the enthalpy to calculate the corresponding multi-objective fitness. In the present work, the similarity in the PXRD pattern of a structure compared to that of a reference ($S$) is chosen as an objective to be minimized, while the enthalpy ($H$) is simultaneously minimized (Equation 1). With $S_s$ and $H_s$ representing the numerical value of the similarity index and enthalpy of structure $s$, respectively, the fitness is defined through the following weighted sum:

$$f_s = w\left(\frac{S_{max} - S_s}{S_{max} - S_{min}}\right) + (1-w)\left(\frac{H_{max} - H_s}{H_{max} - H_{min}}\right) \qquad (1)$$

Here, $w$ is the weight assigned to the PXRD similarity objective, and $A_{min}$ and $A_{max}$ represent the minimum and maximum value of the objective $\{A = S, H\}$ for the pool of structures. The weights of the objective are constrained to be a real number between 0 and 1, and their sum must equal 1 for the calculation of the fitness, $f_s$. This fitness measure, subsequently, is used by XtalOpt to evaluate the suitability of candidate structures for the selection of the parent pool, from which new structures are produced by applying various evolutionary operations, as described more fully in Reference [30].

**PXRD-Assisted Crystal Structure Prediction**

We introduce here the *XtalOpt-VC-GPWDF* coupled technique, whose schematic workflow is illustrated in Figure 1, to conduct PXRD similarity tests during the execution of the crystal structure search. In this multi-objective search, the energy or enthalpy ($H_s$ in Equation 1) is obtained from any external optimizer of periodic systems (herein, we employ the Vienna *ab initio* Simulation Package, VASP, See also Computational Details).[35] The similarity of a structure's simulated PXRD pattern with that of a reference ($S_s$ in Equation 1) is obtained using the newly developed variable-cell Gaussian powder-based similarity index (*VC-GPWDF*),[31] a modified version of de Gelder's similarity index,[36] that ranges from 0 for identical structures to 1 for maximum similarity. Analogous to other methods,[37–39] the similarity is evaluated between a reference PXRD diffractogram, which can be either experimental or computer-generated, and a second diffractogram computed from one of the XtalOpt predicted structures by critic2,[32] which also handles the *VC-GPWDF* similarity index calculation. The reference diffractogram is input as an external list of values containing the 2θ angle of diffraction, and its corresponding relative intensity [2θ; I], and then parsed by critic2. The list does not need to be continuous, nor does it need to cover the whole PXRD diffractogram range. In fact, a short list of a few specific indexed peaks, or just fragments of a PXRD diffractogram, are valid inputs as well. This aspect facilitates the PXRD-assisted crystal structure search, and is particularly useful when multiple phases are present in the sample, or when the noise is large in the experimental data.

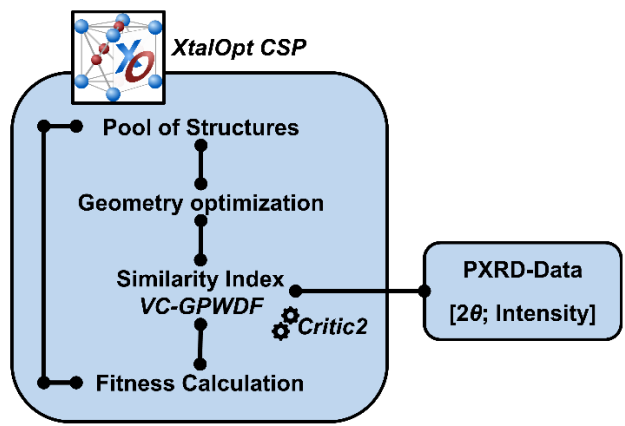

**Figure1. Schematic workflow for conducting the PXRD-assisted crystal structure search with the combined *XtalOpt-VC-GPWDF* method.** The initial pool of structures can be generated internally using the RandSpg algorithm [40], or externally (e.g by PyXtal[41]), with the structures subsequently being read in as "seeds". Local geometry relaxations can be performed with any external code for periodic systems. The similarity index, which is used to calculate the fitness of the offspring, is calculated by critic2 using the optimized geometries of the structures generated by XtalOpt and a set of experimental PXRD data provided by the user in the form of a 2*θ* angle of diffraction and relative intensity list.

The *VC-GPWDF* method is exhaustively described in ref [31]; herein we provide a brief summary. *VC-GPWDF* makes use of a modified version of de Gelder's similarity index (GPWDF), which calculates the overlap between diffraction patterns using a cross-correlation function. In contrast to de Gelder's original approach, GPWDF can be calculated analytically from the list of reflection angles and intensities, which is more efficient and allows for the computation of analytical derivatives with respect to the structural parameters, enabling an easy optimization of the similarity index, i.e., optimizing the strains applied on the structure to minimize the index. *VC-GPWDF* works by performing a global minimization of the GPWDF index over the lattice parameters of the input structure, up to a maximum strain chosen by the user. In the present work, we employed the default values of 10% deformation over the cell length and 5° for the angles. The resulting *VC-GPWDF* index is the smallest GPWDF found among all lattice deformations. In this way, *VC-GPWDF* can overcome differences in diffraction patterns caused by approximations in the computational method chosen to generate the structure, or by thermal and compression effects. Compared to other programs used to perform structure determination using cross-correlation functions, like the Fit with Deviating Lattice parameters (FIDEL),[42] *VC-GPWDF* has two advantages: i) it performs a global optimization, and therefore is unlikely to get caught in a local minimum, and ii) it uses an analytical version of de Gelder's index, enabling fast local optimizations.

Let us now illustrate the power of this new technique for three examples: (1) Brookite, a metastable polymorph of $TiO_2$; (2) sodium ramp-compressed to hundreds of GPa of pressure; and (3) vaterite, a natural polytypic structure of calcium carbonate. Each system will be introduced and discussed in-depth in their specific sections, while the complete computational details are reported at the end of this work.

**Results and Discussion**

**$TiO_2$ – Brookite**

$TiO_2$ naturally exists in three different polymorphs at ambient conditions: Anatase (*I4$_1$/amd*), Brookite (*Pbca*) and Rutile (*P4$_2$/mnm*) with 4, 8 and 2 formula units (FU), respectively, in their conventional unit cells (Figure 2). These polymorphs have been frequently used as benchmarks for CSP methods and related computational models.[19,21,27,40] At the PBE level of theory, we predict Anatase as the ground state, followed by Brookite (ΔE = 13.5 meV/atom) and Rutile (ΔE = 26.7 meV/atom), in-line with previous DFT calculations.[21] Some classic interatomic potentials developed for this system, however, predict different stability orderings.[21]

A typical CSP search performed on $TiO_2$ will almost certainly locate the ground state phase, in this case, Anatase (within the PBE approximation and ensuring that the FUs considered in the search include multiples of 4). While a standard CSP search is also likely to discover metastable Rutile, owing to its high-symmetry and small unit cell, Brookite may be difficult to find due to its metastability and low symmetry. Indeed, in an earlier study, a regular CSP search on TiO$_2$ with 8 FU found Anatase, Rutile and Brookite as the 187th, 559th and 1141st crystals optimized, respectively.[21] A constrained search, where the parent pool was restricted to those structures that possessed an orthorhombic Bravais lattice only, accelerated the discovery of Brookite (as the 345th structure). Another way in which Brookite might be found is by steering an evolutionary algorithm with additional information, such as PXRD data, as we illustrate below.

In the bottom panel of Figure 2, the PXRD pattern of the geometry-optimized structures of the three aforementioned polymorphs of titanium dioxide are plotted. The similarity to Brookite, where the reference PXRD data was generated with Mercury 2022.3.0 [43] in the range from 1° to 120° in 2$\theta$ with a step of 0.1°, from the experimental structure collected by Meagher and Lager at room temperature (ICSD = 36408; lattice parameters in Å: a = 5.138, b = 9.174, c = 5.449)[44] is also provided. Despite the fact that the similarity index (or dissimilarity) is low, 0.01, the DFT-optimized cell parameters are quite different from that of the reference: a = 5.192 Å, b = 9.274 Å, c = 5.509 Å. In fact, if we calculated the similarity index using the DFT-optimized structure directly, as proposed in other methodologies (See introduction), a value of 0.17 would be obtained (to be compared to 0.43 for Anatase and 0.93 for Rutile). Because *VC-GPWDF* performs unit cell deformations, which include varying the DFT-optimized cell parameters during the comparison of the reference and trial structures, it becomes possible to retrieve a nearly zero similarity index for the correct structure. The cell parameters of the relaxed structure post-refinement with *VC-GPWDF* are varied to a = 5.140 Å, b = 9.171 Å and c = 5.447 Å, which almost exactly coincides with the reference structure.

Now that we have described how the *VC-GPWDF* method can modulate a DFT-optimized structure, so that it provides an optimal match with the reference, let us examine the optimal similarity it provides between Brookite and the two higher symmetry polymorphs. The main difference between the diffractogram of orthorhombic Brookite with the one computed for the two tetragonal phases is the number of peaks, which is much larger in the less symmetric case. Moreover, Brookite presents three intense peaks at low angle, two of them very close to each other (25.3°, 25.7° and 30.8° in 2$\theta$ degree, respectively), while in the tetragonal phases, only one high intensity peak is found. One of the reasons why Rutile has a larger similarity index to Brookite than Anatase, is that in Anatase the intense peak is in a range of 2$\theta$ similar to where the first doublet of peaks in Brookite is found, while in Rutile it is at a higher angle (~27°).

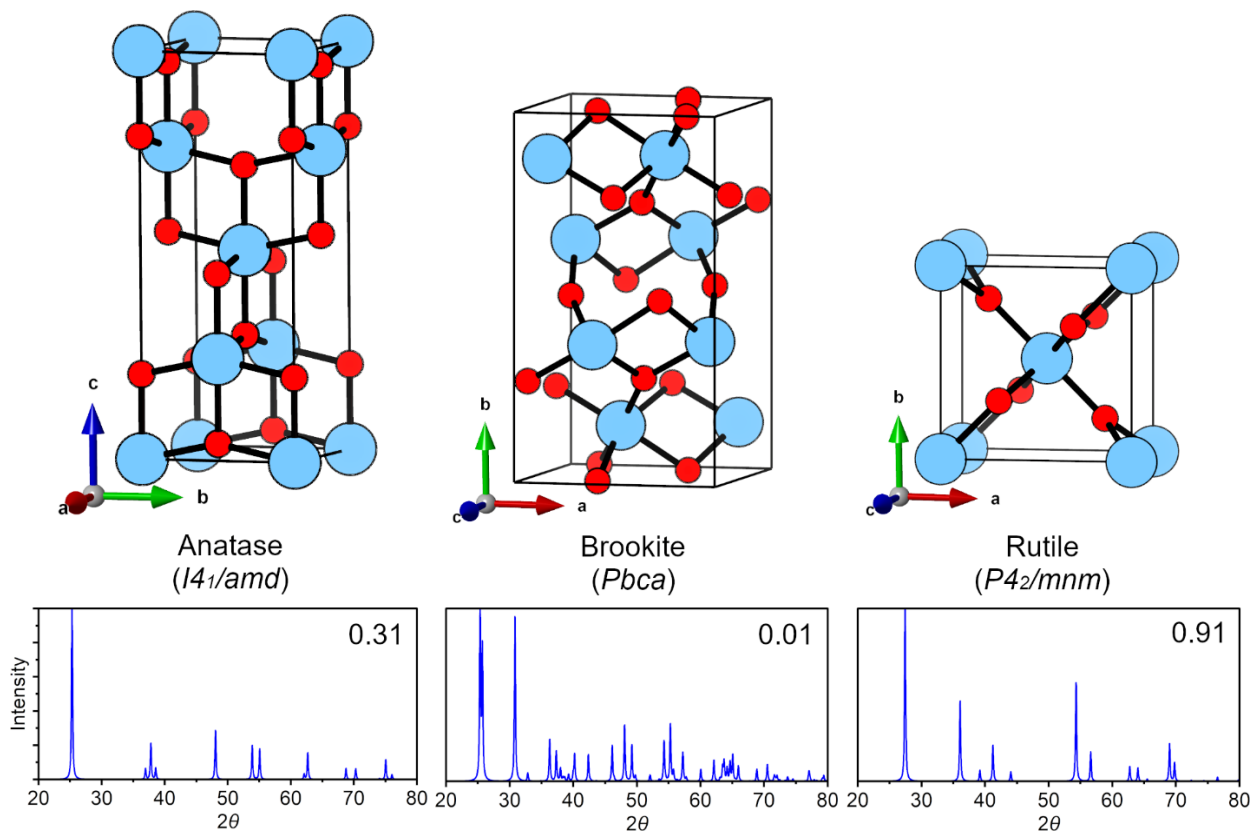

**Figure 2. Polymorphs of TiO$_2$ and their simulated powder X-ray diffraction (PXRD) patterns.** Conventional unit cells of the three natural polymorphs of TiO$_2$ at ambient conditions: Anatase (*I4$_1$/amd*, 4 FU), Brookite (*Pbca*, 8 FU) and Rutile (*P4$_2$/mnm*, 2 FU). Below, we show the PXRD pattern generated from the three phases (λ= 1.54056 Å, which corresponds to the wavelength of Cu *Kα* radiation) and the corresponding similarity index calculated by *VC-GPWDF* using the PXRD-pattern of Brookite as reference. The patterns shown are for the smallest similarity index that can be obtained by varying the unit cell parameters.

Now, let's put the *VC-GPWDF* similarity index in action with XtalOpt to predict the metastable, low symmetry, phase of TiO$_2$, Brookite. To begin, two *single-objective* (classic enthalpy based) CSP runs with 8 FU in the cell (24 atoms) were performed as reference tests generating a total of 1000 structures each (Table 1). In both searches, Brookite was not found, while the more symmetric Anatase and Rutile were generated, in-line with previous studies where 1100+ structures were optimized to find the orthorhombic phase.[21] Coupling XtalOpt with the *VC-GPWDF* algorithm to perform the *multi-objective* search, it was possible to find Brookite in shorter evolutionary searches (See Table 1). However, this success appeared to be sensitive to the fitness weight parameter connected to the PXRD data (Eq. 1). In this test, Brookite was successfully found using $w \geq 0.6$ prompting us to analyze how the choice of the weight influences the fitness of the three polymorphs of TiO$_2$ (Section S1). The fitness is related to the probability that a structure has for being chosen as a parent in the evolutionary search, but there are other factors, including the symmetry and the types of lattices in the initial pool, as well as the random parameters chosen during the course of the CSP, which also influence a structure's discovery. Though the fitness of Brookite was higher than that of Anatase already using a weight of 0.1, this polymorph was not discovered in our short CSP search even when a weight of 0.3 was used. While it is probable that

Brookite could be found in fewer structures than in a regular search (~1100) with this weight, increasing the weight to 0.6 hastens its appearance by more than a factor of two.

**Table 1.** Benchmark tests on $TiO_2$-brookite using *single-* or *multi-objective* crystal structure prediction runs with 8 Formula Units. The number of total structures-per-run (# Structures), the weight assigned to the powder X-ray diffraction similarity objective ($w$), and the output result of the search (if brookite was found or not), are reported. [a]

|  | *single-objective* | | *multi-objective* | | |
|---|---|---|---|---|---|
| Run | 1 | 2 | 1 | 2 | 3 |
| # Structures | 1000 | 1000 | 500 | 500 | 500 |
| $w$ | 0.0 | 0.0 | 0.3 | 0.6 | 0.9 |
| Brookite | No | No | No | Yes | Yes |

[a] If the *Pbca* Brookite phase was fortuitously generated in the initial pool by RandSpg, the run was repeated.

In the past, when the multi-objective PXRD search was not available, it was suggested that constraining the breeding pool to structures whose (sub)lattice was consistent with a particular Bravais lattice or space group (potentially deduced from a diffractogram) could be employed for unveiling the structure of a synthesized compound.[45] However, as shown in the following two examples, the PXRD search proposed herein is preferred since it accounts for possible variations in the crystal lattice (See below, *Na in Ramp-Compression*), and constraining a CSP search with an indexed unit cell might even be counter-productive in some rare cases (See below, *The Tricky Case of Vaterite*).

**Na in Ramp-Compression Experiment**

High-quality, and perhaps already indexed PXRD data can surely increase the success of the *multi-objective* search strategy that we describe above. However, data collected at extreme conditions, such as in dynamic or ramp compression experiments that explore the chemistry of the interiors of planets or high-energy-density quantum matter,[46] often require substantial support from theory for their interpretation. In fact, the data obtained in these experiments is obscured by noisy background, mostly sourced by the hot plasma ablated by the sample target during the laser irradiation,[47] which jeopardizes the indexing of weak reflection's peaks. Therefore, the comparison with PXRD data simulated from theoretical structures is often necessary to identify a phase in shock experiments. Despite this synergistic approach, the structural determination of new phases measured at extreme conditions still remains a great scientific challenge.[48]

A noisy background is not the only challenge in the structural solution from shock and ramp compression experiments. In fact, the kinetics in dynamic compression experiments, together with the uniaxial orientation of the shock front can alter the expected (theoretical) $P - T$ path followed in a phase diagram, greatly diverging from the ideal thermodynamic path at low temperature, and leading to unexpected phase transitions, or even decompositions.[49] Yet, these variables are nothing but additional coordinates of the phase diagram of a compound that must be explored to understand the behavior of matter at extreme conditions.[46] From the theoretical point of view, this means that the system must be simulated with techniques beyond the standard 0 K DFT approximation, including quasi-harmonic phonons to account for the thermal volume expansion,[50] or by performing expensive molecular dynamics simulations.[51] Moreover, the computational reproduction of an anisotropic (*e.g.* uniaxial) compression can be a challenging task even for simple unary systems.[52] Therefore, the possibility to perform volume-cell modulations on-the-fly, and emulate all the effects that contribute to the divergence from an ideal compression experiment, can become useful for a rapid, but meaningful, interpretation of the experimental data. Below, we illustrate the power of this approach on Na, which assumes the iconic *hP*4 insulating electride phase,[53] observed in diamond anvil cells at ~200 GPa.[20]

Studying Na to pressures above 200 GPa, conditions that are accessible only with dynamic compression techniques, is currently of great interest as the findings will address important questions for theory and for condensed matter physics.[54] Therefore, recent laser-driven ramp compression experiments where sodium was squeezed to a nearly 7-fold increase in density at a pressure of 500 GPa (and a temperature of ~1500-3000 K) were performed.[54] In-situ XRD revealed a series of peaks obtained at the highest pressures that could be indexed as the *hP4* phase, but the peaks observed between 242 – 292 GPa were not consistent with *hP4* and were instead interpreted as either the *cI16* structure (previously observed between 108 - 120 GPa[55]) or an *R-3m* phase. A following theoretical study,[56] however, revealed that both *cI16* and *R-3m* were not dynamically stable at the experimentally attained *P-T* conditions. In 0 K CSP searches *hP4* emerges as by far the most stable phase at the pressures attained in experiment, but a subset of seven systems was also found, and computed to be preferred at high temperatures within the quasi-harmonic approximation.[56] Unfortunately, none of their diffraction patterns and densities were fully consistent with the experiments.

To identify the Na structure that was likely created using ramp compression, we carried out a *VC-GPWDF* assisted *multi-objective* CSP search for Na at 315 GPa using the experimental data published in Reference 54, and a weight of 0.7. In this case, the reference list of reflection values is constituted by only six indexed peaks (six [$2\theta$; *I*] pairs), also to avoid contamination from the high noise over the experimental $2\theta$ range (Figure 3a). This search found that the phase producing the best similarity index was actually *hP4*, the expected ground state at these conditions. So why wasn't it previously identified by either experiment or by theory? The answer stems from the severe distortions the structure seems to undergo during the ramp-compression, which cannot be emulated by the standard 0 K DFT optimization, but that is easily revealed by the cell-variation routine in *VC-GPWDF*.

In Figure 3 we illustrate the structure of the *hP4* phase as it emerges from the DFT optimization and post-refinement with *VC-GPWDF* (*hP4**), coupled with the diffraction peaks that these two phases yield overlayed on the XRD data collected by Polsin *et al*.[54] The main structural difference between *hP4* and *hP4*\* is the extra anisotropic compression along the *c*-axis (Figure 3b) estimated to be equal to ~425 GPa (DFT stress value) and compared to the ~370 GPa obtained along the *a*- and *b*-axes, causing the lowered *c/a* ratio and the increased density of the crystal. The subsequent effect of this distortion on the calculated diffraction lines is, on one hand, to move the (101) to higher $2\theta$ angles, and on the other hand to almost merge the reflections from the (102) and the (2-10) planes (Figure 3a), matching with the doublet peaks experimentally observed at ~63°.

In Na-*hP4*\* the density increases up to 6.3 – 6.4 g/cm$^3$, which is relatively high compared to what is expected from 0 K DFT calculations on *hP4*.[20] However, the pressure versus density curve of sodium can deviate quite substantially from the ideal trend, and produce very different results depending on the experimental conditions.[54] For example, at 300 GPa, the density of sodium can be estimated as being ~3.5 g/cm$^3$ following the Sesame principal Hugoniot, or being ~ 5.9 – 6.0 g/cm$^3$ if extrapolated using static compression data from the FCC and BCC phases.[54] Moreover, hP4* is calculated to be 199 meV/atom higher in enthalpy than the fully relaxed hP4 structure (while retaining all real phonons, see Figure S1), which is nonetheless accessible based on the estimated temperature in Polsin's experiment (~ 2200 K).[54] The weak peak at ~42° that was previously indexed as the hP4 (101), was suggested to indicate that multiple phases coexisted, resulting from pressure and temperature gradients present in the sample, in-line with prior interpretations.[54,56]

Na-hP4* is a distorted structure that cannot be obtained with classic DFT geometry optimizations at high-pressure or molecular dynamics simulations, since in both cases, the system would evolve towards the most stable (relaxed) configuration. Instead, Na-hP4* mirrors the effects of anisotropic/uniaxial compression along the c-axis and the thermal expansion, which are extrapolated and accessed thanks to the iterative refinement over the experimental data. Though it is not possible to unequivocally identify the phase observed by Polsin et al.[54] as hP4*, it is worth noting that the PXRD assisted CSP with XtalOpt-VC-GPWDF could access interesting new alternatives for the interpretation of challenging data collected at extreme conditions.[48,57]

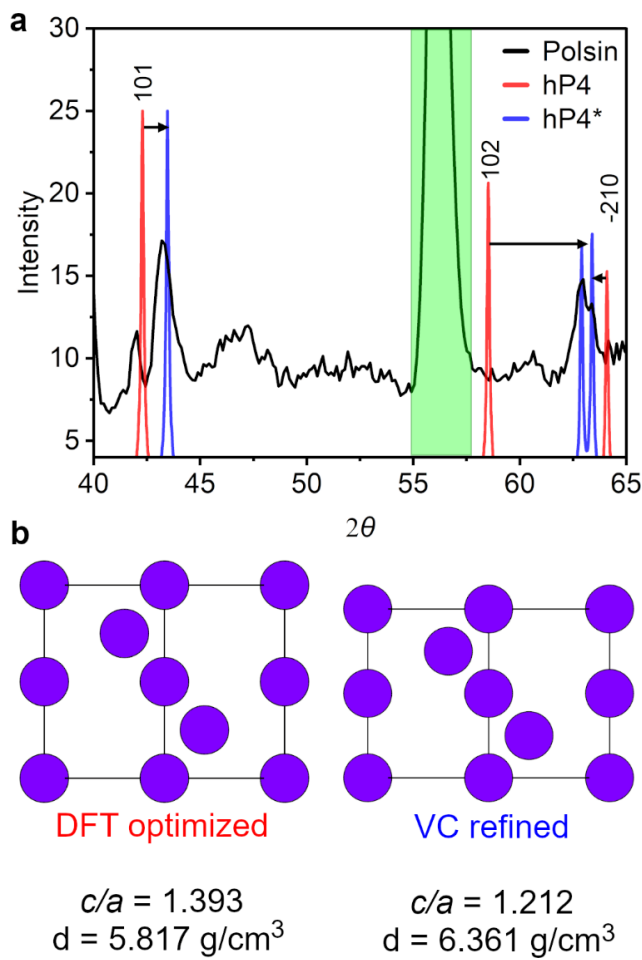

**Figure 3. Simulated XRD patterns and experimental lineouts of Na structures under pressure.** (**a**) Section of the diffraction pattern measured by Polsin et al.[54] (black line) together

with the diffraction lines calculated from the DFT-optimized (red line) and refined (blue line) structures of Na-*hP4* (λ= 1.481 Å, see Ref. 54). The similarity index changed from 0.991 (*hP4*) to 0.086 (*hP4\**) upon refinement. The calibration peak in the diffraction data is shaded in green. (**b**) View of the unit cell of Na-hP4 along the (110) plane, as optimized by DFT (*hP4*), and after the volume-cell refinement with *VC-GPWDF* (*hP4\**).

**The Tricky Case of Vaterite**

Among the biogenic minerals, calcium carbonate ($CaCO_3$) is arguably the most abundant. From the three known anhydrous crystalline polymorphs of $CaCO_3$, calcite, aragonite and vaterite, the latter is the least stable, but still commonly found in nature.[58] Surprisingly, despite the nearly 100-year debate on its crystal structure, an apparently satisfying solution was proposed only very recently.[59] Specifically, it was suggested that vaterite should be regarded as a polytypic structure, a specific type of polymorphism built up by a stacking of almost identical layers, which differ in their stacking sequence. This has made vaterite a very challenging system to solve, even combining several experimental techniques,[59] and almost impossible with computational methods alone. In fact, though vaterite is a relatively simple mineral (composition-wise), it cannot be solved solely with standard CSP methods, even with the possibility of using supercells. Therefore, what can a structure search do to support the solution of such challenging systems? This is what we will try to understand with this last example using our new methodology.

The *multi-objective* evolutionary search coupled with *VC-GPWDF* is obviously limited by the type of PXRD data used. In this case, it was possible to retrieve two extensive lists of peaks [*I*; 2θ] from the studies performed by Le Bail *et al.*[60] and by DuPont *et al.*[61], and two CSP runs using a weight of 0.7 were carried out using one, or the other, as a reference. In Figure 4, we plot the energy difference (relative to calcite) vs. similarity index of the phases output by the CSP runs, focusing on those identified as good matches (similarity index < 0.1).

Using the data indexed by Le Bail *et al.*[60] (Figure 4a), it is not surprising to see that the proposed *Ama2* phase was predicted by our PXRD assisted-CSP as the best match. However, using this list of peaks, our search also found the *Pnma* structure proposed by Meyer,[62] which is ~35 meV/atom more stable than the *Ama2* phase proposed by Le Bail, as well as the *P2₁2₁2₁* phase proposed by DeMichelis[63] (which is isoenergetic to *Pnma*).

Using the second set of data, collected by DuPont *et al.*[61] (Figure 4b), yields different results. The structures proposed previously by LeBail[60] (*Ama2*), Meyer[62] (*Pnma*) and DeMichelis[63] (*P2₁2₁2₁*) were still found. Notice that the similarity index of the recurrent structures changes from one data set to another, but it is consistently very low (< 0.1). Moreover, the two assisted searches have also generated new structures having excellent similarity index and a low energy (only ~10 meV/atom above calcite), and not proposed in past works, which are commented on in Section S4. The most interesting result obtained with DuPont's data is probably the prediction of both the *C2* and *C2/c* structures, previously predicted in another work by DeMichelis *et al.*,[64] which differ by the specific orientation of the carbonate group along the stacking direction of the layers. These two structures are extremely important, since they form the sub-set of phases composing the polytypic crystal structure recently proposed.[59]

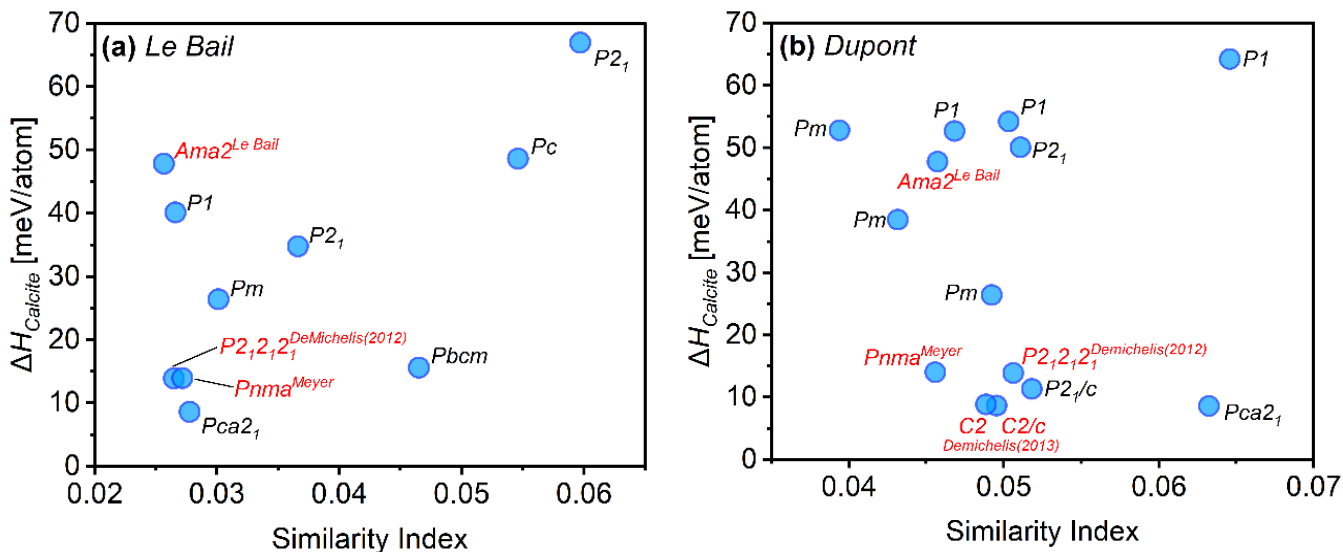

**Figure 4. Plot of the properties of the structures predicted with *XtalOpt-VC-GPWDF* for vaterite.** The relative energies (using calcite as a reference) versus similarity index generated using (**a**) Le Bail's[60] and (**b**) DuPont's[61] experimental powder x-ray diffractograms.

These results show how the coupled *XtalOpt-VC-GPWDF* algorithm can support the solution of complicated crystal structures such as vaterite. Our method was able to generate, almost on-the-fly, most of the crystal structures proposed for vaterite in past theoretical and experimental works, including those forming the polytypic structure, and ranking them by energy and similarity with the experimental PXRD. As we have postulated, even by generating the correct metastable crystal structure, it would have been impossible to thoroughly solve the case of vaterite. However, our new methodology was able to provide all the building blocks necessary to construct the polytypic model that solves the intricate crystal structure of vaterite.

**Discussion**

We have introduced a powder X-ray diffraction-assisted crystal structure prediction method that employs both the enthalpy of a structure and its similarity index, as compared to that of a reference X-ray diffraction pattern, in an equal footing. This technique has been implemented within the open-source evolutionary algorithm code, XtalOpt. The similarity index is calculated using *VC-GPWDF*, a modified version of de Gelder's similarity index, which assesses the overlap between diffraction patterns through a cross-correlation function upon iterative distortions of the unit cells. This similarity index is then used to determine the fitness parameter in XtalOpt's multi-objective global optimization process. Our method is shown to be optimal for identifying metastable phases, facilitating the identification of polymorphs in inorganic samples, and aiding in the analysis of structures distorted by the extreme conditions created in shock and ramp compression experiments. Moreover, it is also effective in identifying challenging structures such as polytypic systems. We believe that the coupled *XtalOpt-VC-GPWDF* tool will be highly beneficial for crystallographers, chemists, materials scientists and geochemists for the solution of challenging structures at ambient and extreme conditions.

## Methods

**Computational Details:** The open-source evolutionary algorithm XtalOpt[30,33] version 13.0 was employed for crystal structure prediction, using the multi-objective fitness measure. The initial generation consisted of random symmetric structures that were created by the RandSpg algorithm,[40] except in the case of $CaCO_3$, where the initial generation was created externally with PyXtal[41] then imported as seeds, using Ca atoms and $CO_3$ trigonal planar units. We believe that this first step could be improved using automated classifications [65], by generating a more accurate initial pool of structures, focusing on the most probable space groups identified by the machine learning engine, a possibility that we will explore in future works. The number of initial structures was equal to 50 in all cases. The number of formula units (FUs) was set equal to 8 in the case of $TiO_2$ to automatically cover the FU of all the natural polymorphs, *i.e.* Anatase (4 FU), Brookite (8 FU) and Rutile (2 FU); 4, 6, 8, 12, 20, 24 and 32 in Na; and 4, 6, 8 and 12 in $CaCO_3$. A sum of the atomic radii scaled by a factor of 0.7 was used to determine the shortest distances allowed between pairs of atoms. Duplicate structures were identified and removed from the breeding pool using the XtalComp algorithm.[66] For the $TiO_2$-brookite test, the total number of generated structures could vary from 500 to 1000 (see Section $TiO_2$ - Brookite). For the tests performed on high-pressure Na and $CaCO_3$, the total number of generated structures per run was equal to 1000. Each structure search followed a multi-step strategy, with three subsequent optimizations with increased level of accuracy, plus a final accurate single point (see below).

Geometry optimizations and electronic structure calculations were performed using Density Functional Theory (DFT) with the Vienna Ab Initio Simulation Package (VASP), version 6.4.2.[35] The PBE[67] exchange-correlation functional was employed. The projector augmented wave (PAW) method[68] was used to treat the core states in combination with a plane-wave basis set with an energy cutoff of 500 eV. The O $2s^22p^4$ (PAW_PBE O_s), Ti $3d^34s^1$ (PAW_PBE Ti), Na $2p^63s^1$ (PAW_PBE Na_pv), Ca $3p^64s^2$ (PAW_PBE Ca_pv) and the C $2s^22p^2$ (PAW_PBE C_s) states were treated explicitly. The *k*-point meshes were generated using the Γ-centered Monkhorst−Pack scheme,[69] and the number of divisions along each reciprocal lattice vector was selected so that the product of this number with the real lattice constant was greater than or equal to a given cutoff. The values of 20, 25 and 30 Å were used for the three subsequent optimization steps in the crystal structure search of $TiO_2$ and $CaCO_3$, then a k-mesh of 50 Å was used for the final single point. In the case of sodium, a k-mesh of 40 Å was used at each optimization step, and one of 50 Å for the final single point. The accuracy of the energy convergence was set to increase from $10^{-3}$ to $10^{-5}$ eV for the optimizations, and to $10^{-6}$ for the final single point on the structures for which the norms of all the forces calculated during the relaxations were smaller than $10^{-3}$. A Gaussian smearing was used at each optimization step, and for each system with a sigma of 0.02 eV. The tetrahedron method was adopted in the last single point.[70]

**Code Availability And Data Availability:** The script and the experimental data used in this work are reported in the Supplementary Information. The methodology will be implemented as a new module in XtalOpt. The refined Na-*hP4\** structure and new *Pca2₁* and *P2₁/c* $CaCO_3$ phases are provided in a separate folder.

**Acknowledgements:** Funding for this research is provided by the Center for Matter at Atomic Pressures (CMAP), a National Science Foundation (NSF) Physics Frontier Center, under Award PHY-2020249 and the NSF award DMR-2119065. Calculations were performed at the Center for Computational Research at SUNY Buffalo (http://hdl.handle.net/10477/79221).

**Author Contributions:** The project was conceived and supervised by E.Z.; S.H. implemented the multi-objective search in the XtalOpt code. S.R. developed the method and performed the calculations and analysis with help from A.O.. S.R. wrote the manuscript, which was edited by E.Z. We thank Xiaoyu Wang for the valuable discussions.

**Competing Interests:** The authors declare no competing interests.

**Additional Information:** Supplementary Information. The online version contains supplementary material available at….